\documentclass[aps,prl,twocolumn,superscriptaddress,showpacs]{revtex4}

\usepackage{graphicx}

\begin{document}


\title{Degenerate Fermi Gases of Ytterbium }


\author{Takeshi Fukuhara}
\affiliation{Department of Physics, Graduate School of Science, Kyoto University, Kyoto 606-8502, Japan}

\author{Yosuke Takasu}
\affiliation{Department of Electronic Science and Engineering, Graduate School of Engineering, Kyoto University, Kyoto 615-8510, Japan}

\author{Mitsutaka Kumakura}
\affiliation{Department of Physics, Graduate School of Science, Kyoto University, Kyoto 606-8502, Japan}
\affiliation{CREST, Japan Science and Technology Agency, Kawaguchi, Saitama 332-0012, Japan}
\affiliation{PRESTO, Japan Science and Technology Agency, Kawaguchi, Saitama 332-0012, Japan}

\author{Yoshiro Takahashi}
\affiliation{Department of Physics, Graduate School of Science, Kyoto University, Kyoto 606-8502, Japan}
\affiliation{CREST, Japan Science and Technology Agency, Kawaguchi, Saitama 332-0012, Japan}

\date{\today}

\begin{abstract}

An evaporative cooling was performed to cool the fermionic $^{173}$Yb atoms in a crossed optical dipole trap. The elastic collision rate, which is important for the evaporation, turns out to be large enough from our study. This large collision rate leads to efficient evaporation and we have successfully cooled the atoms below 0.6 of the Fermi temperature, that is to say, to a quantum degenerate regime. In this regime, a plunge of evaporation efficiency is observed as the result of the Fermi degeneracy.
 
\end{abstract}

\pacs{03.75.Ss, 32.80.Pj, 34.50.-s}


\maketitle
Ultracold fermionic gases have been extensively studied for recent years from the first realization of degenerate Fermi gases of $^{40}$K \cite{DeMarco99}. The degenerate $^{40}$K atoms are also obtained by sympathetic cooling with $^{87}$Rb atoms \cite{Roati02}. Besides, a gas of $^{6}$Li atoms was cooled to degeneracy by sympathetic cooling with bosonic isotope $^{7}$Li  \cite{Truscott01, Schreck01}, $^{23}$Na \cite{Hadzibabic02}, or $^{87}$Rb \cite{Silber05} and a two-spin-state mixture in optical dipole trap \cite{Granade02}. These degenerate atom clouds were applied to investigation into the crossover between Bardeen-Cooper-Schrieffer superfluidity and Bose-Einstein condensation (BEC) of molecules, resulting in remarkable progress \cite{Regal04, Bartenstein04, Zwierlein04, Kinast04}. In systems of the degenerate Fermi gases in an optical lattice, a variety of interesting experiment such as fermionic Bloch oscillations \cite{Roati04} and observation of Fermi surfaces \cite{Kohl05} are repoted. So far, the achieved degenerate Fermi gases are limited only to the two atom species, $^6$Li and $^{40}$K \cite{McNamara06}, because there is no other stable fermionic isotope in the alkali atoms, for which cooling techniques have been established. Fermi degeneracy of other atoms which may have quite different nature from alkali atoms can open out unique possibilities for studies on the ultracold Fermi gases.

\begin{figure}
\includegraphics[width=\linewidth]{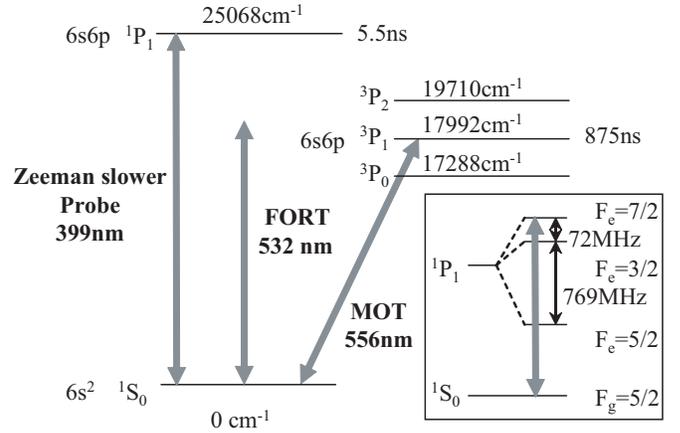}
\caption{Energy levels of Yb atoms. The inset shows the hyperfine structure of $^1$P$_1$ state. The wavelengths of the lasers for cooling, trapping, and probing are also represented. \label{energylevel}}
\end{figure}

Rare-earth atoms of Ytterbium (Yb) have a unique advantage in studying Fermi gases because there exist two stable fermionic isotopes, whose natural abundance is more than ten percents: $^{171}$Yb (14.3 \%) and $^{173}$Yb (16.1 \%) with nuclear spin $I=1/2$ and $I=5/2$, respectively. We can perform experiment on Fermion-Fermion mixture by using these two isotopes. There are also five bosonic isotopes, which provides us with various possibilities of sympathetic cooling pair with favorable collision properties and degenerate Boson-Fermion mixtures. Furthermore, unique features coming from the electron structure, which is similar to the alkaline-earth atoms, enable us to perform unique techniques and important applications. See Fig. \ref{energylevel}. First, no electron spin in the ground state leads to less decoherence caused by stray magnetic field. Second, an ultra-narrow transition of $^1$S$_0-^3$P$_0$ (the natural linewidth $\sim 10$ mHz \cite{Porsev04}), which is an excellent candidate for an atomic clock \cite{Porsev04, Takamoto05} and was precisely observed with an uncertainly of 4.4 kHz \cite{Hoyt05}, enables us to detect a weak energy difference such as pairing gap of Fermi superfluidity. Third, the metastable $^3$P$_2$ state has a large magnetic dipole moment of $3\mu_B$ ($\mu_B$ is the Bohr magneton), resulting in a factor of 9 larger magnetic dipole-dipole interaction than that of alkali atoms, and thus studies on anisotropic superfluidity can be expected. Large mass is also important nature of Yb atoms, which leads to possibilities for interesting heteronuclear molecules with large mass difference \cite{Petrov05}. Because of these unique properties, degenerate Fermi gases of Yb atoms become quite attractive.

In this paper, we report all-optical formation of degenerate fermionic $^{173}$Yb gases. First, laser-cooled atoms are transferred from a magneto-optical trap (MOT) to a vertically and horizontally crossed Far-off-resonance trap (FORT). Reducing the trap depth of the horizontal FORT, we performed evaporative cooling of six-spin-state mixture in the crossed optical dipole trap. At the final stage of evaporation, where the trap depth is a factor of 300 times less than initial one, the temperature of atom clouds is about half of the Fermi temperature. Besides the cooling of $^{173}$Yb atoms, we also investigated a scattering property between different spin components of $^{173}$Yb atoms, because the scattering length is one of the most important parameter in the study of ultracold dilute atoms.

The experiments are performed with essentially the same setup previously used to achieve $^{174}$Yb BEC \cite{Takasu03b}. First, the Yb atoms are decelerated by the Zeeman slower with the strong transition ($^1$S$_0-^1$P$_1$, $F_g=5/2 \to F_e=7/2$, 399 nm) and then cooled and collected by the MOT with the intercombination transition ($^1$S$_0-^3$P$_1$, $F_g=5/2 \to F_e'=7/2$, 556 nm) \cite{Kuwamoto99}, as shown in Fig. \ref{energylevel}. The MOT beam is generated by a dye laser whose frequency is narrowed less than 100 kHz (the linewidth of the $^1$S$_0-^3$P$_1$ transition $\Gamma / 2 \pi$ is 182 kHz) and stabilized by Ultra-Low Expansion cavity, whose stability is typically less than 20 Hz/s. The intensity of each MOT beam is typically 60$I_s$ ($I_s$ is a saturation intensity of $^1$S$_0-^3$P$_1$ transition and $I_s$=0.14 mW/cm$^2$) and the detuning is 7$\Gamma$. The loading time for the MOT is typically 30 seconds, which enables us to prepare $2\times 10^7$ $^{173}$Yb atoms in the MOT.

The intensity of MOT beam is decreased to efficiently transfer the atoms from the MOT to the FORT. In the case of $^{174}$Yb, which has no electronic and nuclear spin in the ground state, the intensity is decreased down to 5$I_s$ in our system, which results in the  temperature for the MOT clouds of about 50 $\mu$K. However, $^{173}$Yb has nuclear spin I=5/2 and effective radiation pressure of MOT beam is smaller due to optical pumping effect \cite{Takasu03a}. So in this experiment of $^{173}$Yb, the intensity is lowered to 25$I_s$. The temperature of $^{173}$Yb in the MOT is 15 $\mu$K and this is low compared with $^{174}$Yb MOT because sub-Doppler cooling mechanism works \cite{Kuwamoto99, Maruyama03}.

The laser-cooled atoms are transferred to a crossed FORT which consists of horizontal and vertical beams. The beams are independently produced by two 10 W diode-pumped solid-state lasers at 532 nm and the $1/e^2$ beam radii at the crossed point are 15 $\mu$m (horizontal beam) and 24 $\mu$m (vertical beam). The initial trap depths are 620 $\mu$K (horizontal) and 30 $\mu$K (vertical), so $2\times 10^6$ atoms at 100 $\mu$K are trapped mainly in the horizontal FORT, whose radial and axial trapping frequencies at full power are 3.6 kHz and 30 Hz, respectively. The radial trapping frequency is measured by parametric resonance methods \cite{Friebel98}. The Fermi temperature of the trapped atoms is about 5 $\mu$K, where we assume that 6 spin-components are equally distributed because $^{173}$Yb atoms in the ground state have no electron spins and the effect from a magnetic field is small.

We measure atom numbers and temperatures using absorption imaging technique. The trapping beams are turned off within less than several hundreds of ns and time-of-flight (TOF) time later the released gas is illuminated by a linearly polarized probe beam pulse resonant with $^1$S$_0-^1$P$_1$ ($F_g=5/2 \to F_e=7/2$) transition propagating along the direction of 0.9-Gauss magnetic field. The intensity of the beam is 0.02 $I_s$, where $I_s$ is the saturation intensity, and the pulse duration is 100 $\mu$s, which is much longer than the absorption cycle of 550 ns at this intensity. It should be noted that the transition rate is different for each spin component and thus a total absorption rate generally depends on the distribution of the spin components. However, it turned out from numerical calculations based on the rate equation that, in the case of our condition, the total absorption rate is the same within 2\% for any initial spin distribution because of the rapid redistribution due to the optical pumping effect of the probe beam, which enables us to determine the total number of atoms exactly, without knowing the population of each component.

To perform the evaporation efficiently, large elastic collision rate is necessary. They are well described as a few partial waves at low temperatures because, when temperatures are much less than the threshold energy $E_{th}(l)$ for a given partial wave $l$ (and there is not a shape resonance \cite {Boesten96, Burke99, Tojo06}), the collisions associated with the partial wave of higher than $l$ are negligible. This threshold energy can be approximately determined by the centrifugal barrier and the van der Waals potential of $C_6/R^6$,  
\begin{equation}
E_{th}(l)=2 \left(\frac{\hbar ^2 l(l+1)}{6 \mu}\right)^{3/2} C_6^{-1/2}
\end{equation}
where $\mu$ is the reduced mass. In the case of fermionic atoms, we can consider only the s-wave scattering between nonidentical fermions at the temperatures much less than the p-wave threshold energy $E_{th}(l=1)$ because the s-wave scattering between identical atoms is zero from Fermi-Dirac statistics. Therefore, the s-wave scattering length between nonidentical atoms plays an important role at the final stage of evaporation toward the degenerate Fermi gases, and so we carried out the cross-dimensional rethermalization technique to deduce the scattering length \cite{Monroe93}.

\begin{figure}
\includegraphics[width=\linewidth]{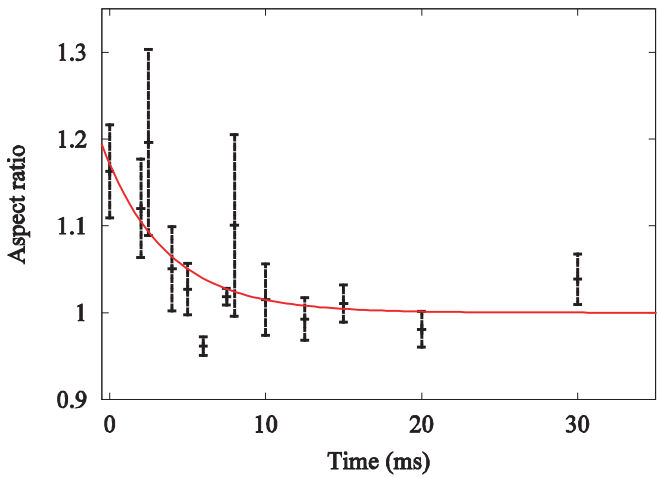}
\caption{Rethermalization process after anisotropic heating of the atom cloud. We caused heating along horizontal direction and observed rethermalization process as the time evolution of temperatures in the horizontal and vertical directions. The rethermalization time $\tau $ is 4 ms.\label{rethermalize}}
\end{figure}

Assuming the coefficient of the van der Waals $C_6=1000$ a.u. \cite{Derevianko}, the p-wave threshold energy is about 60 $\mu$K. Therefore the measurement of rethermalizing process is performed at 6 $\mu$K, where the s-wave scattering is dominant unless there exists a p-wave shape resonance. Typically $1.2\times 10^4$ atoms at 6 $\mu$K are prepared in the crossed FORT and then a 40\% amplitude modulation of 800 Hz, twice the trapping frequency of one horizontal direction, is applied to the power of the vertical FORT during 5ms, which causes heating along that direction. Following the modulation, thermal relaxation is observed as the time evolution of temperatures in the horizontal and vertical directions (Fig. \ref{rethermalize}). An exponential fit to the data extract a rethermalization time $\tau = 4 \pm 1$ ms. To confirm the relaxation is not due to anharmonic mixing, we have also applied this technique with the same configuration of the FORT for $^{176}$Yb atoms, whose scattering length is not large \cite{Takasu06}. The result for the $^{176}$Yb atoms is that the rethermalization time is observed to be longer than 50 ms, which means that the relaxation due to the anharmonicity is not less than 50 ms, therefore the anharmonicity effect is negligible. In the case of $^{173}$Yb atoms, if we assume the atoms are equally populated among the six spin states, the rethermalization time and elastic collision cross section are related by the following equation: $\frac{\alpha}{\tau} = \frac{5}{6} \bar{n} \sigma \bar{v}$, where $\bar{n} = \int n^2(\vec{r}) d^3r/\int n(\vec{r})d^3r$ is the average density and $\bar{v} = 4 \sqrt{(k_B T)/(\pi m)}$ is the mean relative velocity with $k_B$ the Boltzmann constant and m the ytterbium mass. The factor 5/6 comes from the fact that there are no s-wave collisions between identical spin components. The constant $\alpha$ represents how many elastic collisions lead to cross-dimensional rethermalization and turned out to be about 2.7 by the Monte Carlo simulation \cite{Monroe93}. From this analysis, we have deduced the absolute value of the scattering length $|a_s|= 6 \pm 2$, whose error is mainly caused by that of the time $\tau$ and the density $\bar{n}$. This large scattering length allows us to cool the atoms to the quantum degenerate regime.

\begin{figure}
\includegraphics[width=\linewidth]{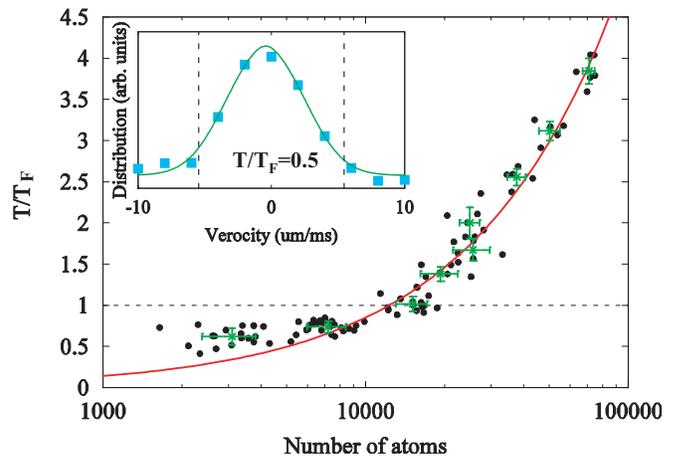}
\caption{Trajectory of evaporative cooling in the crossed FORT. The data of evaporation are shown as solid circle. The solid line (red) shows the evaporation trajectory with $\gamma =2.4$, which is extracted from a fit to the data of the evaporation above the Fermi temperature. Although the evaporation efficiency is nearly constant above the Fermi temperature $T_F$, it decreases below $T_F$ because of the Fermi Pressure and Pauli blocking. The inset shows velocity distribution (solid square) and Thomas-Fermi fit (solid green line) with $T/T_F=0.5$. Two dashed line correspond to the Fermi velocity $v_F=5.5$ $\mu$m/ms. \label{evap}}
\end{figure}

Forced evaporative cooling is carried out by ramping down the intensity of the horizontal beam, while the vertical FORT power is constant. In 2 s, the potential depth of the horizontal FORT is reduced to 30 $\mu$K, which is the same as the one of the vertical FORT, and about $8\times 10^4$ atoms are trapped in the crossed region. The temperature is about 4 $T_{\text{F}}$, where the Fermi temperature $T_{\text{F}}$ is 1.4 $\mu$K at this reduced FORT power. Further decrease of the horizontal FORT power results in the evaporation in the crossed region and the potential depth along the vertical direction, which is considerably affected by the gravity at the final stage of evaporation, is reduced below 2 $\mu$K. A few thousands atoms remain in the trap, whose mean trapping frequency is 450 Hz, and the temperatures are almost half of $T_{\text{F}}$, which is typically 300 nK. At this stage, the absorption images are fitted by a Thomas-Fermi distribution, which is shown in the inset of Fig. \ref{evap}.

The effect of the quantum degeneracy is observed as the decrease of the cooling efficiency \cite{DeMarco99}. Figure \ref{evap} shows the number of atoms N and the temperature renormalized by the Fermi temperature $\bar{T}=T/T_F$ at each stage of the evaporation. A line represents a trajectory of evaporation with the cooling efficiency $\gamma = -\frac{d(\ln \rho)}{d (\ln N)}= 3 \frac{d(\ln \bar{T})}{d (\ln N)}=2.4$, which is extracted from a fit to the data above the Fermi temperature. Below the Fermi temperature, the efficiency falls due to the Fermi pressure and the Pauli blocking \cite{DeMarco99}.  

One strategy for the decrease in the elastic collision rate at the final stage of the evaporative cooling is to take evaporation time longer. However, we observed a rapid atom loss of several hundreds of ms, which is an obstacle to the improvement of cooling efficiency. The rate of background gas collisions and photon scattering by the FORT beams is considerably small to explain the loss rate. We consider the observed rapid atom loss is due to three-body recombination. This is consistent with the fact that the three-body loss rate is proportional to $a_s^4$, which becomes large for $^{173}$Yb whose scattering length is found to be large. If it is the case, we can reduce this rapid loss by preparing two-spin mixture using optical pumping, allowing us to obtain further deeply degenerate gases.

In conclusion, we have succeeded in cooling fermionic $^{173}$Yb atoms down to 0.6 $T_{\text{F}}$, where the quantum degeneracy is observed as the decrease in the evaporation efficiency. We have also studied the elastic collision rate, revealing that the elastic collision is large enough to perform efficient evaporation and the scattering length is about 6 nm assuming that the spin components are equally distributed.

It is interesting to apply degenerate gases of $^{173}$Yb for researches on Fermi superfluidity. Although a technique of Feshbach resonances, which plays an important role in those studies, can not be used for Yb because of no hyperfine state in the ground state, there exists an alternative method to control the scattering length using photoassociations \cite{Fedichev96}, so-called optical Feshbach resonances, which is predicted to effectively work for the narrow linewidth transition such as the $^1$S$_0-^3$P$_1$ transitions \cite{Ciurylo05}. The ultra-narrow $^1$S$_0-^3$P$_0$ transition also provides a desirable probe for the superfluid pairing gap. In the future, degenerate Fermi gases of Yb atoms with these techniques can be a powerful tool toward outstanding goals in this field such as high $T_{\text{c}}$ superfluidity in optical lattices \cite{Hofstetter02}.

We acknowledge T. Yabuzaki for helpful discussion. This work was partially supported by Grant-in-Aid for Scientific Research of JSPS (18043013, 18204035), SCOPE-S, and 21st Century COE "Center for Diversity and Universality in Physics" from MEXT of Japan. T. F. would like to acknowledge support from Yoshida Scholarship Foundation, and Y. Takasu from JSPS.


\begin{thebibliography}{99}
\bibitem{DeMarco99}B. DeMarco and D. S. Jin, Science \textbf{285}, 1703 (1999)
\bibitem{Roati02}G. Roati, F. Riboli, G. Modugno, and M. Inguscio, Phys. Rev. Lett. \textbf{89}, 150403 (2002)
\bibitem{Truscott01}A. G. Truscott, K. E. Strecker, W. I. McAlexander, G. B. Partridge, and R. G. Hulet, Science \textbf{291}, 2570 (2001)
\bibitem{Schreck01}F. Schreck, L. Khaykovich, K. L. Corwin, G. Ferrari, T. Bourdel, J. Cubizolles, and C. Salomon, Phys. Rev. Lett. \textbf{87}, 080403 (2001)
\bibitem{Hadzibabic02}Z. Hadzibabic, C. A. Stan, K. Dieckmann, S. Gupta, M. W. Zwierlein, A. G\"{o}rlitz, and W. Ketterle, Phys. Rev. Lett. \textbf{88}, 160401 (2002)
\bibitem{Silber05}C. Silber, S. G\"{u}nther, C. Marzok, B. Deh, P. W. Courteille, and C. Zimmermann, Phys. Rev. Lett. \textbf{95}, 170408 (2005)
\bibitem{Granade02}S. R. Granade, M. E. Gehm, K. M. O'Hara, and J. E. Thomas, Phys. Rev. Lett. \textbf{88}, 120405 (2002)
\bibitem{Regal04}C. A. Regal, M. Greiner, and D. S. Jin, Phys. Rev. Lett. \textbf{92}, 040403 (2004)
\bibitem{Bartenstein04}M. Bartenstein, A. Altmeyer, S. Riedl, S. Jochim, C. Chin, J. H. Denschlag, and R. Grimm, Phys. Rev. Lett. \textbf{92}, 120401 (2004)
\bibitem{Zwierlein04}M. W. Zwierlein, C. A. Stan, C. H. Schunck, S. M. F. Raupach, A. J. Kerman, and W. Ketterle, Phys. Rev. Lett. \textbf{92}, 120403 (2004)
\bibitem{Kinast04}J. Kinast, S. L. Hemmer, M. E. Gehm, A. Turlapov, and J. E. Thomas, Phys. Rev. Lett. \textbf{92}, 150402 (2004)
\bibitem{Roati04}G. Roati, E. de Mirandes, F. Ferlaino, H. Ott, G. Modugno, and M. Inguscio, Phys. Rev. Lett. \textbf{92}, 230402 (2004)
\bibitem{Kohl05}M. K\"{o}hl, H. Moritz, T. St\"{o}ferle, K. G\"{u}nter, and T. Esslinger, Phys. Rev. Lett. \textbf{94}, 080403 (2005)
\bibitem{McNamara06}Recently, the Fermi degeneracy of metastable $^3$He atoms was achieved by sympathetic cooling with $^4$He. J. M. McNamara \textit{et. al.}, available at http://arxiv.org/abs/cond-mat/0606638. 
\bibitem{Porsev04}S. G. Porsev and A. Derevianko, Phys. Rev. A \textbf{69}, 042506 (2004)
\bibitem{Takamoto05}M. Takamoto, F.-L. Hong, R. Higashi, and H. Katori, Nature \textbf{435}, 321 (2005)
\bibitem{Hoyt05}C. W. Hoyt, Z. W. Barber, C. W. Oates, T. M. Fortier, S. A. Diddams, and L. Hollberg, Phys. Rev. Lett. \textbf{95}, 083003 (2005)
\bibitem{Petrov05}D. S. Petrov, C. Salomon, and G V Shlyapnikov, J. Phys. B \textbf{38}, S645 (2005)
\bibitem{Takasu03b}Y. Takasu, K. Maki, K. Komori, T. Takano, K. Honda, M. Kumakura, T. Yabuzaki, and Y. Takahashi, Phys. Rev. Lett. \textbf{91}, 040404 (2003)
\bibitem{Kuwamoto99}T. Kuwamoto, K. Honda, Y. Takahashi, and T. Yabuzaki, Phys. Rev. A \textbf{60}, R745 (1999)
\bibitem{Takasu03a}Y. Takasu, K. Honda, K. Komori, T. Kuwamoto, M. Kumakura, Y. Takahashi, and T. Yabuzaki, Phys. Rev. Lett. \textbf{90}, 023003 (2003)
\bibitem{Maruyama03}R. Maruyama, R. H. Wynar, M. V. Romalis, A. Andalkar, M. D. Swallows, C. E. Pearson, and E. N. Fortson, Phys. Rev. A \textbf{68}, 011403(R) (2003)
\bibitem{Friebel98}S. Friebel, C. D'Andrea, J. Walz, M. Weitz, and T. W. H\"{a}nsch, Phys. Rev. A \textbf{57}, R20 (1998)
\bibitem{Boesten96}H. M. J. M. Boesten, C. C. Tsai, B. J. Verhaar, and D. J. Heinzen, Phys. Rev. Lett. \textbf{77}, 5194 (1996)
\bibitem{Burke99}J. P. Burke, Jr., C. H. Greene, J. L. Bohn, H. Wang, P. L. Gould, and W. C. Stwalley, Phys. Rev. A \textbf{60}, 4417 (1999)
\bibitem{Tojo06}S. Tojo, M. Kitagawa, K. Enomoto, Y. Kato, Y. Takasu, M. Kumakura, and Y. Takahashi, Phys. Rev. Lett. \textbf{96}, 153201 (2006)
\bibitem{Monroe93}C. R. Monroe, E. A. Cornell, C. A. Sackett, C. J. Myatt, and C. E. Wieman, Phys. Rev. Lett. \textbf{70}, 414 (1993)
\bibitem{Derevianko}A. Derevianko (private communication); V. Pal'chikov {\it ibid}; K. P. Geetha {\it ibid}; The value of $C_6$ of Yb$_2$ can be estimated by $ \langle^{1}S_0||D||^{1\!}P_1\rangle^4/(3 dE)$, with $D$ the dipole moment and $dE$ the energy separation. Depending on the calculation, the range of $C_6$ takes from 1000 a.u. to 2000 a.u.
\bibitem{Takasu06}Y. Takasu, T. Fukuhara, M. Kitagawa, M. Kumakura, and Y. Takahashi, Laser Phys. \textbf{16}, 713 (2006)
\bibitem{Fedichev96}P. O. Fedichev, Y. Kagan, G. V. Shlyapnikov, and J. T. M. Walraven, Phys. Rev. Lett. \textbf{77}, 2913 (1996)
\bibitem{Ciurylo05}R. Ciurylo, E. Tiesinga, and P. S. Julienne, Phys. Rev. A \textbf{71}, 030701(R) (2005)
\bibitem{Hofstetter02}W. Hofstetter, J. I. Cirac, P. Zoller, E. Demler, and M. D. Lukin, Phys. Rev. Lett. \textbf{89}, 220407 (2002)

\end{thebibliography}
\end{document}